\documentclass{elsart}
\usepackage{graphicx}
\usepackage{amsmath}
\usepackage{bbm}
\usepackage{psfrag}
\usepackage{graphicx}
\usepackage{latexsym}

\newcommand{\bG}{\bf G}
\newcommand{\bg}{\bf g}
\newcommand{\bC}{\bf C}
\newcommand{\bc}{\bf c}
\newcommand{\bX}{\bf X}

\begin{document}
\begin{frontmatter}

\title{\LARGE\bf Signal and Noise in Correlation Matrix}

\author{Z. Burda}, 
\author{A. G\"orlich}, 
\author{A. Jarosz} and  
\author{J. Jurkiewicz}

\address{M. Smoluchowski Institute of Physics,
Jagellonian University, Reymonta 4, 30-059~Krak\'ow, Poland}

\begin{abstract}
Using random matrix technique we determine 
an exact relation between the eigenvalue 
spectrum of the covariance matrix and of its
estimator. This relation can be used in
practice to compute eigenvalue invariants
of the covariance (correlation) matrix.
Results can be applied 
in various problems where one experimentally
estimates correlations in  a system with many degrees of freedom,
like for instance those in statistical physics, 
lattice measurements of field theory, genetics, quantitative finance
and other applications of multivariate statistics.
\end{abstract}


\end{frontmatter}

Statistical systems with 
many degrees of freedom appear in numerous research areas.
One of the most fundamental
issues in studies of such systems is the determination
of correlations.
In practice, one encounters frequently the following
situation: one samples the system many times by
carrying out independent measurements. For each sample
one estimates values of the elements of the covariance matrix,
and then takes the average over a set of samples. 
The statistical uncertainty 
of the average of individual elements of the matrix
generically decreases with the number
of independent measurements $T$ as $\sim 1/\sqrt{T}$.
There are $N(N+1)/2$ independent elements of the correlation
matrix for a system with $N$ degrees of freedom. Thus,
naively, the total uncertainty encoded
in the correlation matrix may be expected to be
proportional to $N(N+1)/2$ and $1/\sqrt{T}$, and therefore
to be large for 'non-local' quantities which depend 
on many elements of the correlation matrix 
even if one performs a large number of
measurements, of the order of the number 
of degrees of freedom in the system. 
It turns out that this naive expectation is far
from true. Such 'non-local' quantities
occur, in particular, in the eigenvalue analysis of the correlation
matrix~\cite{SILVER}. A question which we address in this paper 
is how the spectrum of the experimentally measured 
covariance (two-point correlation) matrix is related to the spectrum
of the genuine correlation matrix for the system.

To be specific, consider a statistical system consisting of
$N$ real degrees of freedom $x_i$, $i=1,\dots,N$ 
with a stationary probability distribution:
\begin{equation}
p\left(x_1,\dots,x_N\right) \prod_{n=1}^N d x_n
\label{measure}
\end{equation}
such that 
\begin{equation}
\int x_i \; p(x_1,\dots,x_N) \prod_{n=1}^N d x_n = 0 \quad \forall i \ .
\end{equation}
The covariance matrix for the system is defined as
\begin{equation}
C_{ij} = \int x_i x_j \; p(x_1,\dots x_N) \prod_{n=1}^N d x_n \ .
\label{C}
\end{equation}
Further, assume that the system belongs to the Gaussian universality
class.  Under this assumption the
probability distribution can be  approximated  by
\begin{equation}
p(x_1,\dots ,x_N) \prod_{n=1}^N d x_n =
\left[ (2\pi)^N {\rm det} {\bf C}\right]^{-1/2} 
\exp \left( -\frac{1}{2} \sum_{ij} x_i C^{-1}_{ij} x_j \right) \ 
\prod_{n=1}^N d x_n
\label{gauss}
\end{equation} 
where $C_{ij}$ is a covariance matrix (\ref{C}) of the system.
By construction it is a symmetric, positive-definite matrix.
In fact, for a wide class of models, the Gaussian approximation
well describes the large $N$ behavior 
of the system as a consequence of the central limit theorem.
Deviations from the Gaussian behavior can 
result either from the presence of fat (heavy) tails in 
the probability distribution 
or from collective excitations of many degrees of freedom. 
None of these  effects will be discussed here.

Experimentally, the correlation matrix is computed as follows.
One performs a series of $T$ independent measurements. 
Assume $T > N$. The measured values $x_n$ form a rectangular
$N\times T$ matrix $\bX$ with elements $X_{nt}$,
where $X_{nt}$ is the measured value of the $n^{\rm th}$ degree of
freedom $x_n$ in the $t^{\rm th}$ experiment $t=1,\dots,T$.
The experimental correlation matrix is
computed using the following estimator
\begin{equation}
c_{ij} = \frac{1}{T} \sum_{t=1}^T X_{it} X_{jt} = 
\frac{1}{T} \left\{ \bX \bX^\tau \right\}_{ij}
\label{CE}
\end{equation}
where $\bX^\tau$ is the transpose of $\bX$.
We expect that for $T\rightarrow\infty$
the estimated values $c_{ij}$ will 
approach the elements $C_{ij}$.
More precisely, if the measurements are independent, 
the probability distribution of measuring a matrix $\bX$ of 
values $X_{nt}$ is a product of probabilities for individual
measurements
\begin{equation}
P({\bX}) D {\bX} = \prod_{t=1}^T \left( p(X_{1t},\dots X_{Nt})
\prod_{n=1}^N d X_{nt} \right) 
\end{equation}
where 
\begin{equation}
D {\bX} = \prod_{n,t=1}^{N,T} d X_{nt} \ .
\end{equation}
In particular, for the Gaussian approximation
\begin{eqnarray}
P({\bX}) D {\bX} & = & {\mathcal N} \exp \left( -\frac{1}{2} 
\sum_{t=1}^T X_{it} C^{-1}_{ij} X_{jt} \right) 
\prod_{n,t=1}^{N,T} d X_{nt} \nonumber \\ & = &
{\mathcal N} \exp \left( -\frac{1}{2} {\rm Tr}\; {\bX}^\tau {\bC}^{-1} {\bX} \right)  
D {\bX}
\label{Pgauss}
\end{eqnarray}
where ${\mathcal N}$ is a normalization factor which ensures that
$\int P({\bX}) D {\bX} = 1$. In this particular 
case ${\mathcal N} = [(2\pi)^N {\rm det} {\bC}]^{-T/2}$.
All averages over measured values $X_{nt}$ are calculated with
this probability measure. We shall denote these averages
by $\langle \dots \rangle$. In particular we see that
\begin{equation}
\langle X_{it} X_{jt'} \rangle = C_{ij} \delta_{tt'} \ . 
\label{propagator}
\end{equation}
This relation reflects the assumed absence of correlations
between measurements.
In general, if measurements are correlated,
the right-hand side of the last equation can be expressed by
a matrix ${\mathcal C}_{it,jt'}$ in  double indices.

After these introductory remarks, we come  
 back to the problem of  relating   the spectrum
of the covariance matrix $\bC$ to the spectrum of its estimator
$\bc$. 
We denote the eigenvalues of the matrix $\bC$ by $\Lambda_n$
$n=1,\dots, N$. For a given set of eigenvalues we
can calculate matrix invariants, 
like for example the spectral moments 
\begin{equation}
M_k = \frac{1}{N} {\rm Tr}\; {\bC}^k = 
 \frac{1}{N} \sum_{n=1}^N \Lambda_n^k = 
\int d \Lambda \; \rho_0(\Lambda) \Lambda^k
\end{equation}
where the density of eigenvalues $\rho_0(\Lambda)$
is defined as
\begin{equation}
\rho_0(\Lambda) = 
\frac{1}{N} \sum_{n=1}^N \delta\left(\Lambda - \Lambda_n\right) \ .
\end{equation}
The question is how  these quantities are related to the analogous quantities
defined for the estimator of the correlation matrix $\bc$
\begin{equation}
m_k = \frac{1}{N} \langle {\rm Tr}\; {\bc}^k \rangle = 
\int d \lambda \rho(\lambda) \lambda^k
\end{equation}
where the eigenvalue density of the matrix estimator is
\begin{equation}
\rho(\lambda) = 
\frac{1}{N} \left\langle
\sum_{n=1}^N \delta\left(\lambda - \lambda_n\right)\right\rangle \ .
\end{equation}
We expect that the dependence of the estimated spectrum 
$\rho(\lambda)$ and the genuine spectrum $\rho_0(\Lambda)$
should be controlled by $T$ and $N$. 
Indeed, as we shall see later, it turns out that for $N \to \infty$
this dependence is governed by the parameter $r=N/T$, which we assume 
to be finite. 

In order to derive the relation between the spectral properties of the
covariance matrix and its estimator it is convenient
to define resolvents: 
\begin{equation}
{\bG}(Z) =  \big( Z \mathbbm{1}_N - {\bC} \big)^{-1}
\label{GC}
\end{equation}
and 
\begin{equation}
{\bg}(z) = 
 \left\langle \big( 
z \mathbbm{1}_N - {\bc} \big)^{-1} \right\rangle =
 \left\langle 
\big( z \mathbbm{1}_N - \frac{1}{T} {\bX} {\bX}^\tau \big)^{-1} \right\rangle
\label{tG}
\end{equation}
where $Z$ and $z$ are  complex variables. The symbol
$\mathbbm{1}_N$ stands for the $N \! \times \! N$ unit matrix.
Expanding the resolvents in $1/Z$ (or $1/z$) one sees that   
they can be interpreted as
generating functions for the moments
\begin{equation}
M(Z)=\frac{1}{N}{\rm Tr} [Z{\bG}(Z)] - 1  = 
\sum_{k=1}^\infty \frac{1}{Z^k} M_k 
\label{mz1}
\end{equation}
and
\begin{equation}
m(z)=\frac{1}{N}{\rm Tr} [z{\bg}(z)] - 1 = 
\sum_{k=1}^\infty \frac{1}{z^k} m_k \ . 
\label{mz2}
\end{equation}
{}From the relation  between $M(Z)$ and $m(z)$
one can determine 
the corresponding relation between the
eigenvalue spectra $\rho_0(\Lambda)$ and $\rho(\lambda)$.
Indeed,  taking the imaginary part of ${\rm Tr}\; {\bg}(z)/N$ 
(and ${\rm Tr}\; {\bG}(Z)/N$) 
for $z = \lambda + i 0^+$ (or $Z  = \Lambda + i 0^+$), 
where 
$\lambda$ is real, we can directly
calculate the eigenvalue densities 
$\rho(\lambda)$ (and $\rho_0(\Lambda)$):
\begin{equation}
\rho(\lambda) = -\frac{1}{\pi} {\rm Im}
\,\,\frac{1}{N}{\rm Tr}\; {\bg}(\lambda + i0^+) 
\label{rho_cut}
\end{equation}
as follows from the standard relation for distributions:
$(x+i0^+)^{-1} = {\rm PV} x^{-1} - i \pi \delta(x)$, where PV stands
for principal value. 

The fundamental relation between the generating functions 
(\ref{mz1}) and (\ref{mz2}) is derived in the Appendix by means
of a diagrammatic technique \cite{fz} for 
calculating integrals (\ref{tG}) with the Gaussian measure
(\ref{Pgauss}). 
The large $N$ limit corresponds to the planar limit in which only 
planar diagrams contribute. This significantly simplifies considerations
and allows one to write down closed formulae for the resolvents. 

This fundamental relation between the generating 
functions (\ref{mz1}) and (\ref{mz2}) reads 
\begin{equation}
m(z) = M(Z)  
\label{main}
\end{equation}
where the complex number $Z$ is related to $z$ by the
conformal map (\ref{ztz}):
\begin{equation}
Z = \frac{z}{1 + r m(z)}
\label{ma1}
\end{equation}
or equivalently, if we invert the last relation for $z = z(Z)$,
as: 
\begin{equation}
z = Z \left(1 + r M(Z)\right).
\label{ma2}
\end{equation}

The equations (\ref{main},\ref{ma1}) were already announced in our
earlier work~\cite{ECONO}. They can for example be used to 
compute moments of the genuine correlation function $\bC$ 
from the experimentally measured moments of the estimator $\bc$.
Indeed, combining (\ref{main}) and (\ref{ma1}) we obtain the
following equation:  
\begin{equation}
m(z) = M\left(\frac{z}{1 + rm(z)}\right)
\end{equation}
which gives a compact relation between moments
$m_k$ and $M_k$:
\begin{equation}
\sum_{k=1}^\infty \frac{m_k}{z^k} = 
\sum_{k=1}^\infty \frac{M_k}{z^k}
\left(1 + r \sum_{l=1} \frac{m_l}{z^l}\right)^k  
\end{equation}
from which we can recursively express $m_k$ by 
$M_l$, $l=1,\dots,k$
\begin{equation}
\begin{array}{ll}
m_1 & = M_1 \\
m_2 & = M_2 + r M_1^2 \\
m_3 & = M_3 + 3 r M_1 M_2 + r^2 M_1^3 \\
\dots & \\
\end{array}
\end{equation}
or inversely: $M_k$ by $m_l$, $l=1,\dots,k$:
\begin{equation}
\begin{array}{ll}
M_1 & = m_1 \\
M_2 & = m_2 - r m_1^2 \\
M_3 & = m_3 - 3 r m_1 m_2 + 2 r^2 m_1^3 \\
\dots & . \\
\end{array}
\label{detrm}
\end{equation}
Let us observe that for $r<1$ the functions $M(Z)$ and $m(z)$ can also be
expanded around $z=Z=0$. In this case 
\begin{equation}
M(Z)=-\sum_{k=0}^{\infty}Z^k M_{-k},
\end{equation}
where
\begin{equation}
M_{-k}=\frac{1}{N}{\rm Tr}{\bC}^{-k} \ .
\end{equation}
Similarly
\begin{equation}
m(z)=-\sum_{k=0}^{\infty}z^k m_{-k},
\end{equation}
where
\begin{equation}
m_{-k} = \frac{1}{N} \langle {\rm Tr}\; {\bc}^{-k} \rangle.  
\end{equation}
Using the same manipulation as before we obtain
\begin{equation}
\sum_{k=1}^\infty M_{-k}Z^k = 
\sum_{k=1}^\infty m_{-k}Z^k
(1-r-r\sum_{l=1}^\infty M_{-l}Z^l)^k
\end{equation}
and hence:
\begin{equation}
\begin{array}{ll}
M_{-1} & = (1-r)m_{-1} \\
M_{-2} & = (1-r)^2 m_{-2} -r(1-r) m_{-1}^2 \\
M_{-3} & = (1-r)^3 m_{-3} -r(1-r)^2 m_{-1} m_{-2} - r^2(1-r) m_{-1}^3 \\
\dots & .\\
\end{array}
\end{equation}
The relations between moments 
can be used directly in practical applications to clean
the spectrum of the correlation matrix.
Before discussing this let us make a comment.
The formulae (\ref{main}) and (\ref{ma1}) encode
full information about the relation between
the eigenvalue spectrum $\rho_0(\Lambda)$ and
$\rho(\lambda)$ for a given $r$. In particular,
if one knows the spectrum $\rho_0(\Lambda)$ 
of the correlation matrix $\bC$ one can exactly determine for a given
$r$ the shape of the spectrum $\rho(\lambda)$ of the estimator
dressed by statistical fluctuations.  
One does it as follows. From the eigenvalue spectrum 
$\rho_0(\Lambda)$ one deduces an explicit form 
of the function $M(Z)$ and of the right hand side of the equation (\ref{ma2}).
Inverting the equation (\ref{ma2}) for $Z$ one finds the dependence
$Z = Z(z)$. Inserting it to the equation
(\ref{main}) one determines the function $m(z)$. 
Taking the imaginary part along the cuts of the map $m(z)$
on the real axis (\ref{rho_cut})
one eventually finds $\rho(\lambda)$.
One can easily write a numerical program which realizes this
procedure. In few cases the solution is possible analytically.
Let us shortly discuss them.

Consider the correlation matrix
$\bC$ whose spectrum is given by a sequence of degenerate eigenvalues $\mu_i$, 
$i=1,\dots,K$ with degeneracies $n_i$. Consequently, defining $p_i=n_i/N$,
$\sum_i p_i = 1$, we have
\begin{equation}
M(Z)=\sum_{i=1}^K \frac{p_i\mu_i}{Z-\mu_i} \ .
\label{MZmu}
\end{equation}
This form is particularly simple to discuss. One should  however
keep in mind that the relations 
(\ref{main},\ref{ma1}) remain valid also in a more general case, for
instance, when in the limit $N\to \infty$ the spectrum 
of $\rho_0(\Lambda)$ is not a sum of delta functions but approaches
some continuous distribution. 
The map (\ref{ma2}) now reads 
\begin{equation}
z = Z\left(1   + r\sum_{i=1}^K \frac{p_i\mu_i}{Z-\mu_i}\right).
\end{equation}
Clearly, if we solve this equation for $Z=Z(z )$ 
we obtain a multi-valued function, except in the case $r=0$ 
for which we have a simple relation $z=Z$. 
The ``physical'' Riemann sheet 
of the map $Z=Z(z)$
is singled out by the
condition $Z\to z$ for $z \to \infty$.
On this sheet the complex $z$-plane is mapped on a part of the 
$Z$ plane without a simply or multiply connected
region surrounding the poles at $Z=\mu_i$. 

As an illustration, let us consider the simplest case, 
where $K=1$. In this case we have only one eigenvalue $\mu_1 = \mu$ and
$p_1 =1$ and $M(Z)=\mu/(Z-\mu)$. The
map (\ref{ma1}) has the following form
\begin{equation}
z = Z  +r \frac{Z \mu}{Z-\mu} \ .
\label{mma1}
\end{equation}
If one rewrites the right-hand side of this equation using
polar coordinates $(R,\phi)$ around the pole: 
$Z - \mu = R e^{i\phi}$:
\begin{equation}
z = R e^{i\phi} + \frac{r\mu^2}{R} e^{-i\phi} + \mu(1+r)
\end{equation}
one can see that the equation is invariant under the  ``duality'' 
transformation:
\begin{equation}
R \longleftrightarrow \frac{r \mu^2}{R} \ , \quad 
\phi \longleftrightarrow -\phi
\label{duality}
\end{equation}
which maps the inside of the circle
$|Z-\mu| = \mu \sqrt{r}$ onto the outside and vice versa.
Obviously, the outside  corresponds to the ``physical'' Riemann sheet
of the inverse map $Z=Z(z)$ 
\begin{equation}
Z = \frac{1}{2}\left((1-r)\mu+z +\sqrt{(z-\mu_+)(z-\mu_-)}\right),
\label{mex1}
\end{equation}
since in this region $Z \sim z$ for $z \to \infty$.
The two constants in the last equation are 
$\mu_\pm = \mu(1 \pm \sqrt{r})^2$. 
Along the cut $\mu_- < z < \mu_+$ on the real axis the map $Z=Z(z)$ 
becomes complex and ambiguous: it has a phase (sign) ambiguity 
which is related to the
fact that the cut is mapped into the limiting 
circle where the two Riemann sheets meet.

From (\ref{main}) we can easily find the generating function
$m(z)$ and then from (\ref{rho_cut}) the spectral density 
of the correlation matrix $\bc$:
\begin{equation}
\rho(\lambda)=\frac{1}{2\pi \mu r}
\frac{\sqrt{(\mu_+-\lambda)(\lambda -\mu_-)}}{\lambda}.
\label{cut2}
\end{equation}
This is a well known result
in random matrix theory~\cite{GAUSS2} for the 
spectral distribution of the Wishart ensemble.
It is interesting to interpret this result as 
a statistical smearing of the initial
spectral density $\rho_0(\Lambda)$ given by the delta function 
localized at $\mu$ into a wide peak $\rho(\lambda)$  
supported by the cut $[\mu_-, \mu_+]$, due to a finite series 
of measurements. The larger $r$ the larger is the width of the
resulting distribution $\rho(\lambda)$. For $r=1$ the
lower limit of the distribution is at $\mu_-=0$ 
signaling the appearance of zero modes 
in the matrix $\bc$. One can show by considering 
anti-Wishart matrices that above the limiting value $r=1$
the zero mode sector of $\bc$ grows with $r$ reflecting 
an increasing indeterminacy of the spectrum of the
covariance matrix $\bC$ when the underlying statistical 
sample becomes too short.

\begin{figure}

\begin{center}
\psfrag{r=0.01}{\footnotesize $r=0.01$}
\psfrag{r=0.115}{\footnotesize $r=0.115$}
\psfrag{r=0.3}{\footnotesize $r=0.3$}

\psfrag{lam}{\footnotesize $\lambda$}
\psfrag{rho}{\footnotesize $\rho(\lambda)$}

\includegraphics[width=7cm]{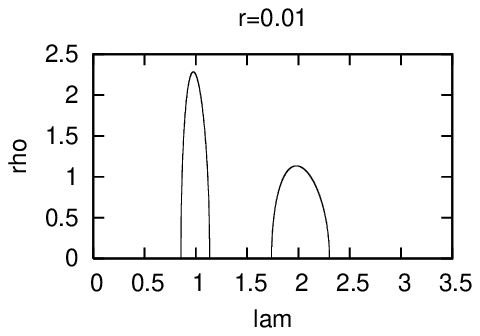}
\includegraphics[width=7cm]{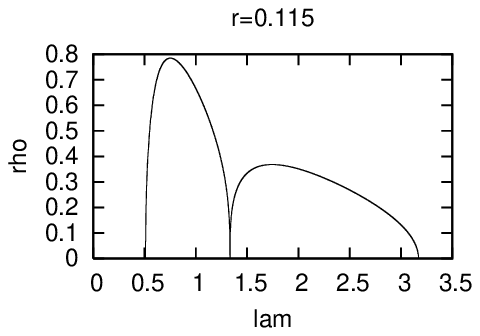}
\includegraphics[width=7cm]{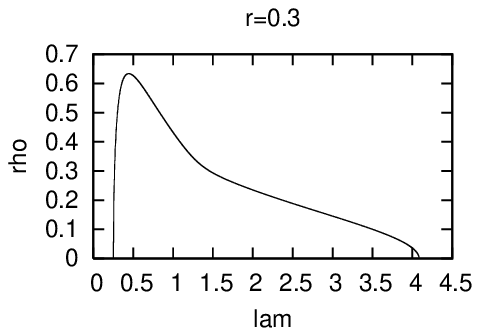}

\end{center}
\caption{The figures represent spectra of the eigenvalue
distributions $\rho(\lambda)$
of the experimental 
correlation matrix measured in a series of measurements
for $r=0.01, 0.115, 0.3$, respectively.
The underlying correlation matrix has two eigenvalues
$\mu_1=1$ and $\mu_2=2$ with the weights $p_1=p_2=0.5$.
At the critical value $r=r_c=0.115$ 
(eq. \ref{rc}) the spectrum splits.
The spectral densities are calculated analytically.
\label{2evt}}
\end{figure}

As a second example we consider the case, where the genuine covariance
matrix $C$ has two different eigenvalues $\mu_1,~\mu_2$ with 
relative weights $p_1,~p_2$, $p_1+p_2=1$. In this case we can also
find an explicit form of the map $Z(z)$ solving the corresponding cubic 
(Cardano) equation.  Depending on the parameters $\mu_i,~p_i$ 
the map $Z(z)$ has one or two 
cuts on the real axis on the $z$ plane, which means that the corresponding 
eigenvalue distribution $\rho(\lambda)$ has a support on one or two  
intervals (fig.\ref{2evt}). 
It is a simple exercise
to find the critical value $r_c$ of $r$ at which 
a single cut solution splits into a two-cut one:
\begin{equation}
r_c = 
\frac{(\mu_2-\mu_1)^2}{\left((p_1 \mu_1^2)^{1/3}+(p_2 \mu_2^2)^{1/3}\right)^3}.
\label{rc}
\end{equation}
For instance, this formula gives $r_c \approx 0.01$ 
if the correlation matrix
$\bC$ has two eigenvalues $\mu_1=1$ and $\mu_2=1.1$ 
and the corresponding weights $p_1=p_2=1/2$. Thus in this
case, to observe a bimodal signal in the measured spectrum 
one has to perform $T$ measurements with $T$ of order $100 N$. 
If the gap $\Delta \mu$ between the eigenvalues $\mu_1$ and $\mu_2$
is larger the binomial structure is visible
already for smaller $T$. As an example we show
in fig.\ref{2evt} 
typical shapes of the distribution
for $\mu_1=1$, $\mu_2=2$, $p_1=p_2=1/2$ for
$r<r_c$, $r=r_c$ and $r>r_c$.
The shape of the spectral functions for $r$ in the range from
$1$ to, say, $0.2-0.3$
resembles that for a single eigenvalue. When $r$ approaches $r_c$
it starts to deviate from this shape  
developing a double peak 
structure which eventually splits into two separate parts at $r=r_c$.
If $r$ is further decreased 
the two peaks get narrower. They eventually entirely
localize at $\mu_1=1$ and $\mu_2=2$ for $r=0$, that is
for an infinitely long sample for which the spectrum
of the underlying correlation matrix is recovered.

The method can be straightforwardly generalized from 
$K=1,2$ to arbitrary $K$,  $\mu_1,\dots, \mu_K$ with 
$\sum p_i =1$, although only the $K=3$ case is solvable 
analytically (quartic Ferrari equation). In other cases one can 
use a numerical implementation of the general
procedure, which we described before, to
determine the shape of the spectrum of the
estimator $\rho(\lambda)$ from any
given distribution $\rho_0(\Lambda)$ and for any $r$. 

In practice one is however interested in the opposite problem
that is in the determination of the spectrum $\rho_0(\Lambda)$
of the genuine correlation matrix $\bC$
from the distribution of the measured eigenvalues. 
This is a difficult problem for the following reason. 
Having one sample of $T$ measurements of
a system with $N$ degrees of freedom one has
as a result only $N$ eigenvalues of the correlation matrix. From
$N$ random numbers it is impossible to reconstruct accurately
the exact form of the underlying distribution 
$\rho(\lambda)$ according to which
they are distributed. Thus the input function $\rho(\lambda)$
for the procedure leading from $\rho(\lambda)$ to $\rho_0(\Lambda)$ 
has a large statistical uncertainty. In effect
the output function $\rho_0(\Lambda)$ is not well controlled.

In some applications one is not interested in the exact
shape of the spectral function $\rho_0(\Lambda)$ but in its
moments. In such cases one can directly make use of the relations
(\ref{detrm})
between the measured moments $m_k$ and the moments of the original
distribution $\rho_0(\Lambda)$ to determine the latter ones.

Moreover, the relations (\ref{detrm}) between the moments may
in some cases significantly improve the determination of the eigenvalues
$\Lambda$ of the original distribution. Assume that
on top of the purely statistical information obtained
by independent measurements of a system consisting of $N$-degrees
of freedom we have at our disposal some additional non-statistical
knowledge about the system. For example, assume
that we know that degrees of freedom can be grouped into 
$K$ sectors of degenerate independent constituents. Each sector is 
represented by an eigenvalue $\mu_k$ and a degeneracy $n_k$, or 
equivalently by the fraction $p_k = n_k/N$ of eigenvectors 
related to this eigenvalue.
In other words, we assume that the resolvent $M(Z)$ 
(\ref{mz1}) of the correlation function $\bC$ is given by 
the formula (\ref{MZmu}) with a specific $K$. 
Thus the problem of determining eigenvalues $\Lambda_i$ 
is reduced to the problem of determining parameters 
$\mu_k$, $p_k$. If $K\ll N$ the problem has much less unknowns.
This non-statistical knowledge of the system 
can be used as follows.  We 
can  explicitly express the moments $M_k$ by the yet unknown
parameters $p_j,\mu_j$. We denote the corresponding functions by
$M_k^{th}(p_j,\mu_j)$. On the other hand we can measure 
experimentally the moments $m_k$ and from the relations
(\ref{detrm}) we can find the corresponding values which we
denote by $M^{exp}_k(m_j)$. Using the jack-knife procedure
we can also estimate the statistical errors $\Delta_k$
of $M^{exp}_k$'s. Minimizing $\chi^2(p_j,\mu_j)$:
\begin{equation}
\chi^2 = \sum_{k=1}^L 
\left( \frac{M_k^{th}(p_j,\mu_j) - M^{exp}_k}{\Delta_k}\right)^2
\label{chi2}
\end{equation}
we can eventually find optimal values of the parameters
$p_j,\mu_j$. Let us make a few comments. Obviously, $L$ must
be equal or larger than the number of unknown parameters. If it
is equal then the minimization of $\chi^2$ amounts to solving
equations $M_k^{th}(p_j,\mu_j) = M_k^{exp}$. In practice, if
possible, $L$ should be taken larger than the number of free
parameters. In this case the weights $1/\Delta_k^2$ 
in $\chi^2$ take care of the gradually 
decreasing importance of the higher moments which are
usually estimated with larger errors.

Let us illustrate the method at work
using the following exercise as an example. 
The exercise has two parts. First we generate $T\times N$ matrix
$\bX$ from the Gaussian distribution (\ref{Pgauss}) with 
a given covariance matrix $\bC$ which has $K$
eigenvalues $\mu_k$ with the weights $p_k$, $k=1,\dots, K$.
Then we treat the parameters $\mu_k, p_k$ as unknown,
and experimentally determine their
optimal values using the method of  moments.
In the end we compare the measured values with those
used in the generator. We repeat this procedure many times
to estimate the error one makes, when one estimates the
spectrum in a single measurement.  

\begin{figure}
\begin{center}

\psfrag{lam}{\footnotesize $\lambda, \Lambda$}
\psfrag{rho}{\footnotesize $\rho(\lambda)$}
\psfrag{Rho}{\footnotesize $\rho_0(\Lambda)$}

\includegraphics{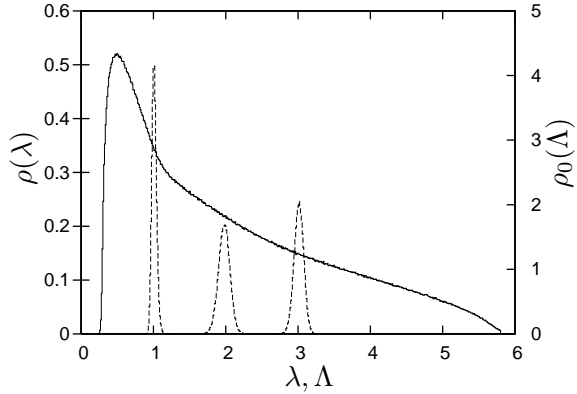}

\end{center}
\caption{The figure 
represents eigenvalue distributions measured
in $n=10^5$ measurements for $N=100$, $T=333$ 
(solid line) and the cleaned spectrum 
using additional information  
that there are three equally probable eigenvalues (dashed line);
The peaks in the right
figure localize around correct values $\mu_1=1$, $\mu_2=2$ and 
$\mu_3=3$ despite the asymmetry parameter is $r=0.3$.
\label{3evt}}
\end{figure}

As a first example we consider the case $K=2$, $\mu_1=1$,
$\mu_2=2$ and $p_1=p_2=0.5$ as before and
with $N=100$ and $r=N/T=0.3$. 
\begin{figure}
\begin{center}

\psfrag{lam}{\footnotesize $\lambda$}
\psfrag{rho}{\footnotesize $\rho(\lambda)$}
\psfrag{Lam}{\footnotesize $\Lambda$}
\psfrag{Rho}{\footnotesize $\rho_0(\Lambda)$}

\includegraphics[width=8cm]{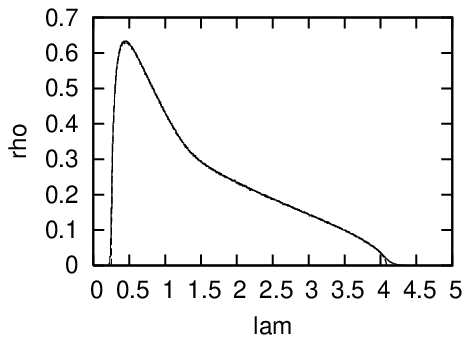}
\includegraphics[width=8cm]{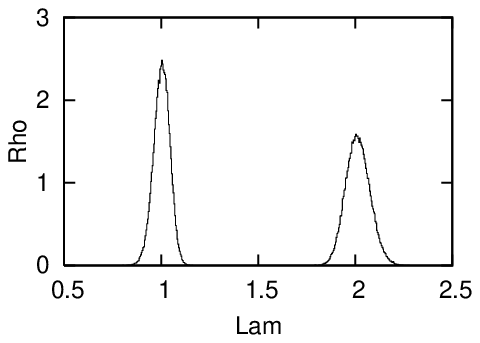}
\includegraphics[width=8cm]{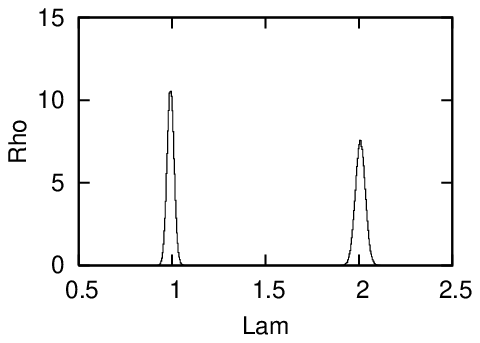}

\end{center}
\caption{The figures 
represent eigenvalue distributions measured
in $n=10^5$ measurements for $N=100$, $T=333$ 
(upper); the cleaned spectrum 
using the additional information 
that there are exactly two eigenvalues (middle);
and using the information
that there are two equally probable eigenvalues 
in the correlation matrix (lower). The peaks in the last two
figures localize around correct values $\mu_1=1$ and $\mu_2=2$ 
despite the number of measurements is relatively small: $N/T=r=0.3$.
\label{2eve}}
\end{figure}
The theoretical curve of the distribution
$\rho(\lambda)$ is shown again in fig. \ref{2eve}
where it is compared with the experimentally determined
distribution. The experimental curve is obtained by averaging
over $n=10^5$ independent experiments. In each experiment
the matrix ${\bX}$ was generated and the eigenvalues of 
$\bc$ were calculated. Thus the resulting experimental
histogram was constructed out of $n \cdot N=10^7$ eigenvalues.
One sees that the experimental curve fits very well to the 
theoretically predicted. Actually, one cannot distinguish
by bare eye the two curves on the first plot in fig. \ref{2eve} .
If we knew only this curve
we would not be able to conclude from it that the underlying 
covariance matrix has only two eigenvalues. Now assume that we know 
that there are exactly two eigenvalues and use the method of moments 
to compute them.  We have to measure at least $L=3$ moments
$m_l$, $l=1,2,3$ in order to determine by minimizing $\chi^2$
(\ref{chi2}) the three unknown parameters $\mu_1$, $\mu_2$, $p_1$.
The resulting spectrum $\rho_0(\lambda)$ averaged over $n$ experiments
is shown in the next plot in fig. \ref{2eve}. 
The spectrum clearly shows the double
peak structure with the correct localization of peaks
around $\mu_1=1$ and $\mu_2=2$. The areas under the peaks are approximately 
equal which means that $p_1\approx p_2 \approx 0.5$, as expected. 
Assume that on top of the information that there are exactly two eigenvalues
we additionally know that both are equally probable $p_1=p_2=0.5$. 
In effect, we have only two unknown parameters 
$\mu_1$ and $\mu_2$. As we see in fig. \ref{2eve} 
this additional information leads to the further sharpening of
the resulting spectrum around the
expected values $\mu_1=1$ and $\mu_2=2$. 
Another example, for $K=3$ and $\mu_1=1$ $\mu_2=2$ and
$\mu_3=3$ is shown in fig. \ref{3evt}.
The method of moments works very well
and allows us to localize the positions of
eigenvalues of the underlying covariance matrix $\bC$.
It is clear that additional information considerably
reduces the error of the estimate.

To summarize: In the limit $N \to \infty$  
we give an exact relation  between the
spectrum of the true correlation matrix
between the degrees of freedom in a statistical
system and the averaged spectrum of the natural estimator of this correlation
matrix used in a statistical sampling. The latter depends on the
parameter $r=N/T$. For finite $r$ the spectrum  of the estimator
is smeared and hides behind the statistical noise the information
about the positions of eigenvalues of the exact correlation matrix.
The averaged spectrum is very similar to that of a spectrum obtained
from a single matrix $c_{ij}$ (\ref{CE}) used in practical applications.
The exact relation can be formulated as an infinite sequence of linear 
relations between the moments
of the true correlation matrix and of its averaged estimator. In principle
both sets are equivalent. In practice, the estimator is known only
as a single realization of the measuring process and the estimated
moments are known only approximately. The relations between the moments
can be nevertheless used 
to determine the spectrum of the true correlation
matrix from a single estimator. 
We illustrate the accuracy of the method of moments on
simple examples.  
The method of moments can be used in practical applications. To give an
example, the spectral analysis of the covariance matrix
plays the central role in the portfolio assessment. 
The knowledge of the covariance matrix is crucial for the optimal 
asset allocation. Typically one constructs the
covariance matrix for $N$ of order $500$ assets' price changes
(e.g. SP500 data) and estimates it using a sample of say four 
years of a daily data. In this case $T$ is of order $1000$ 
and hence $r=0.5$. As we have learned,
in this case one almost entirely looses the signal coming from the 
eigenvalues of the genuine covariance matrix. 
On the other hand, it is known that in practice
there are only few collective sectors on the market. These
sectors are represented by $K\ll N$ significant eigenvectors 
and related eigenvalues. One can use the  method of moments
to localize them \cite{bj}. 

\medskip

\noindent
{\bf Acknowledgments}

\medskip

\noindent
This work was partially supported by the EC IHP Grant
No. HPRN-CT-1999-000161, by the Polish State Committee for
Scientific Research (KBN) grants
2P03B 09622 (2002-2004) and 2P03B-08225 (2003-2006),
and by EU IST Center of Excellence "COPIRA".

\medskip

\noindent
{\bf Note added} 

\medskip

\noindent
After completing our work we became aware of the following papers where
some of the issues raised in our work are also discussed:
V.A. Marchenko and L.A. Pastur, Math. USSR-Sb {\bf 1} (1967) 457;
J.W. Silverstein and Z.D. Bai, J. Multivariate Anal. {\bf 54} (1995)
175; F. Lillo and R.N. Mantegna, cond-mat/0305546.

\medskip

\noindent
{\bf Appendix}

\medskip

We apply a diagrammatic method \cite{fz}
to calculate the integral (\ref{tG}). Expanding the 
expression under the average (\ref{tG}) we obtain a series
of polynomials in ${\bX}$, which can be  integrated term by term 
for the Gaussian measure $P({\bX}) D{\bX}$ (\ref{Pgauss}). 
It is convenient to write the
matrix ${\bc}$ as a product of three matrices
${\bc} = {\bX} \frac{\mathbbm{1}_T}{T} {\bX}^\tau$, the 
first of which is an $N \! \times \! T$ matrix, the second 
is $T \! \times \! T$ , and the third $T \! \times \! N$.
This reveals a sandwich structure of the terms
of the expansion:
\begin{eqnarray}
{\bg}(z) & = &
\left\langle \frac{1}{z - {\bc}} \right\rangle = 
\left\langle \sum_{k \geq 0} \frac{{\bc}^k}{z^{k+1}} \right\rangle 
\label{gz} \\
& = & \left\langle \frac{\mathbbm{1}_N}{z} \quad + 
\frac{\mathbbm{1}_N}{z} {\bX} \frac{\mathbbm{1}_T}{T} {\bX}^\tau 
\frac{\mathbbm{1}_N}{z} \ + 
\frac{\mathbbm{1}_N}{z} {\bX} \frac{\mathbbm{1}_T}{T} {\bX}^\tau 
\frac{\mathbbm{1}_N}{z} {\bX} \frac{\mathbbm{1}_T}{T} {\bX}^\tau 
\frac{\mathbbm{1}_N}{z}
+ \dots \right\rangle \nonumber \\
& = & \left\langle
\psfrag{k}{$\dots$}
\psfrag{+}{$+$}
\includegraphics[width=10cm]{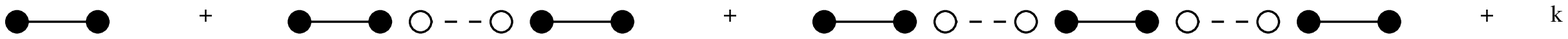}
\qquad \right\rangle
\nonumber 
\end{eqnarray}
The following graphical representation is used in the last
equation. The matrix $\bX$ has two types of indices $X_{it}$,
$i=1,\dots, N$ and $t=1,\dots, T$. We shall call the first
an index of the $N$-type, and the second an index of the $T$-type.
Indices of the $N$-type are drawn as filled circles and 
of the $T$-type as empty circles, a matrix $\mathbbm{1}_N/z$ 
having both indices of the $N$-type is drawn as a solid line 
joining two filled circles corresponding to these indices, 
while a matrix $\mathbbm{1}_T/T$ as a dashed line which joins
two empty circles representing two indices of the $T$-type
of the matrix. An insertion of ${\bX}$ is drawn as
an ordered pair of neighboring circles consisting 
of a filled circle followed by an open one,
while an insertion of ${\bX}^\tau$
as an ordered pair of an open and a filled circle.

The $\bX$ insertions have
to be integrated with the Gaussian measure $P({\bX})D{\bX}$. 
The terms $\frac{\mathbbm{1}_N}{z}$ and $\frac{\mathbbm{1}_T}{T}$ 
are constant from the point of view of the integration.
The integration of the ${\bX}$ insertions amounts to calculating 
the correlation functions
$\langle X_{i_1t_1} \dots X_{i_{2k}t_{2k}} \rangle$. 
Using the Wick theorem we consecutively calculate
$2k$-point correlation functions: 
the two-point correlation function
\begin{equation}
\langle X_{it} X_{jt'} \rangle = C_{ij}\delta_{tt'}
\label{2p}
\end{equation}
the four-point  
\begin{eqnarray}
\nonumber
\langle X_{i_1t_1} X_{i_2t_2} X_{i_3t_3} X_{i_4t_4} \rangle & = &
\langle X_{i_1t_1} X_{i_2t_2}\rangle \langle X_{i_3t_3} X_{i_4t_4} \rangle \\
& + & \langle X_{i_1t_1} X_{i_3t_3}\rangle 
\langle X_{i_2t_2} X_{i_4t_4} \rangle 
\label{4p} \\
& + & \langle X_{i_1t_1} X_{i_4t_4}\rangle 
\langle X_{i_2t_2} X_{i_3t_3} \rangle 
\nonumber
\end{eqnarray}
and higher correlation functions. All odd correlation functions vanish.
All even functions
can be expressed as sums of all distinct products of two-point
functions
 given by (\ref{2p}).
This observation leads to a particularly simple 
graphical representation: a two-point correlation function 
is drawn as a double line  consisting of a solid line
which connects two filled circles, and of a dashed line which connects
two empty circles representing indices of ${\bX}$ and $\bX^\tau$.
We associate the contribution 
$C_{ij}$ with the solid line and $\delta_{tt'}$ with the
dashed line. We will draw the double
lines as semi-circles. Using this convention 
the four-point correlation function can be drawn as
\begin{equation}
\begin{array}{l}
\langle X_{i_1t_1} X_{i_2t_2} X_{i_3t_3} X_{i_4t_4} \rangle = \\
\psfrag{i1}{\scriptsize $i_1$}
\psfrag{i2}{\scriptsize $i_2$}
\psfrag{i3}{\scriptsize $i_3$}
\psfrag{i4}{\scriptsize $i_4$}
\psfrag{t1}{\scriptsize $t_1$}
\psfrag{t2}{\scriptsize $t_2$}
\psfrag{t3}{\scriptsize $t_3$}
\psfrag{t4}{\scriptsize $t_4$}
\psfrag{+}{\large $+$}
\psfrag{=}{\large $=$}
\psfrag{dots}{\large $\dots$}
\includegraphics[width=11cm]{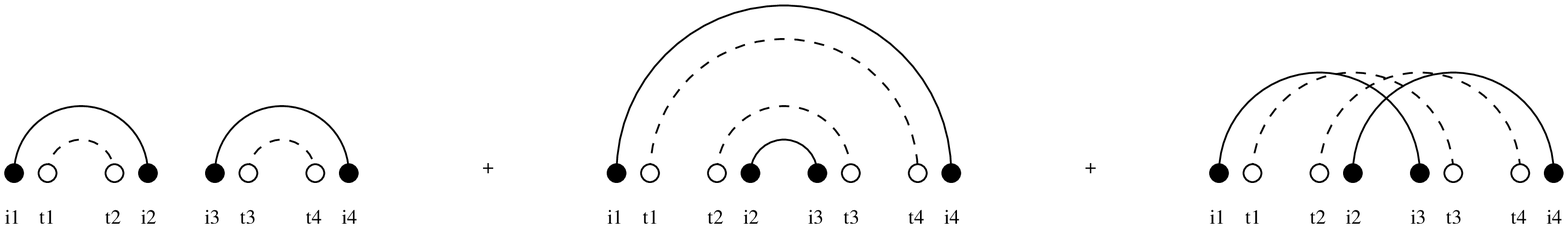}
\label{xxxx}
\end{array}
\end{equation}
Putting together equations (\ref{gz}) and (\ref{xxxx}) we eventually
arrive at the diagrammatic representation of ${\bg}(z)$:
\begin{equation}
\psfrag{+}{$+$}
\psfrag{=}{$=$}
\psfrag{g}{${\bg}$}
\psfrag{k}{$\dots$}
\includegraphics[width=12cm]{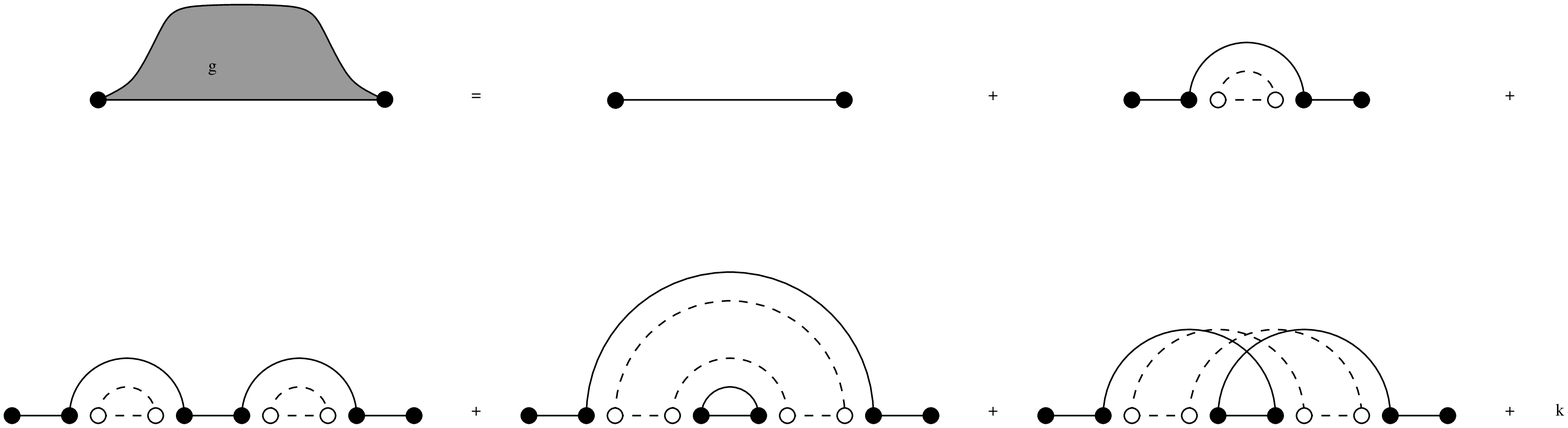}
\label{fgraph}
\end{equation}
We are interested in the limit $N\rightarrow \infty$, $r=N/T={\rm const}$.
In this limit only planar graphs contribute to ${\bg}(z)$, while
non-planar diagrams are suppressed by  factors proportional to powers of
$1/N \rightarrow 0$. This property follows from a simple counting argument.
Each closed internal line of a diagram gives a factor $N$ (or $T$) 
coming from taking a trace.
Thus a closed solid line contributes a factor proportional to $N$, 
and a closed dashed line to $T$. Because we are interested in the
limit of a constant ratio $N/T$,  $T$ is proportional to $N$.
Each horizontal dashed line introduces a  factor 
$1/T \sim 1/N$. For a planar diagram
the number of closed loops is equal to the number of 
dashed horizontal lines and thus the powers of $N$ and of $1/N$
cancel giving a contribution of order unity,
while for a non-planar diagram, like for instance the last one
displayed in  (\ref{fgraph}), 
the power of $1/N$ coming from the number of dashed horizontal lines
is larger than the power of $N$ coming from
the number of closed loops, and thus effectively the contribution
of the diagram has at least one unbalanced factor $1/N$ which makes
it vanish in the limit $N\rightarrow \infty$, $r=N/T={\rm const}$.

From here on we neglect the non-planar diagrams. In parallel to
the generating function ${\bg}(z)$ one can define a
generating function:
\begin{equation}
{\bg}_*(z) =
\left\langle \frac{1}{T\mathbbm{1}_T 
- \frac{1}{z}{\bX}^\tau {\bX}} \right\rangle 
\end{equation}
which contains all diagrams of the same shape as in
(\ref{fgraph}) but with dashed and solid lines replaced. 
It is convenient to additionally introduce 
a class of one-line irreducible diagrams
having the property that they cannot be divided
into two separate diagrams by cutting a single horizontal line.
Denote the sum of one-line irreducible diagrams with two
external vertices of the $N$-type by ${\bf \Sigma}(z)$, and with
two external vertices of the $T$-type by ${\bf \Sigma}_*(z)$.
The sum of diagrams ${\bg}(z)$ can be expressed as a geometric series 
of the ${\bf \Sigma}(z)$ blocks connected by
solid horizontal lines and similarly the sum of diagrams
${\bg}_*(z)$ by the ${\bf \Sigma}_*(z)$ blocks connected by
dashed lines:
\begin{eqnarray}
{\bg}(z) & = &
\left\langle \frac{1}{z - \frac{1}{T} {\bX}{\bX_{\tau}}} \right\rangle 
 =  \frac{\mathbbm{1}_N}{z} + 
\frac{\mathbbm{1}_N}{z} {\bf \Sigma}(z) \frac{\mathbbm{1}_N}{z}  +
\frac{\mathbbm{1}_N}{z} {\bf \Sigma}(z) \frac{\mathbbm{1}_N}{z} {\bf \Sigma}(z)
\frac{\mathbbm{1}_N}{z} + \dots \nonumber \\ & = &
\frac{1}{z \mathbbm{1}_N - {\bf \Sigma}(z)} 
\label{GS} \\
{\bg}_*(z) & = &
\left\langle \frac{1}{T\mathbbm{1}_T 
- \frac{1}{z}{\bX}^\tau {\bX}} \right\rangle 
  = \frac{1}{T\mathbbm{1}_T - {\bf \Sigma}_*(z)}
\end{eqnarray}
The last equations can be represented graphically as:
\begin{center}
\psfrag{s}{{\small ${\bf \Sigma}$}}
\psfrag{g}{{\small $\bg$}}
\psfrag{k}{{\small $\dots$}}
\includegraphics[width=13cm]{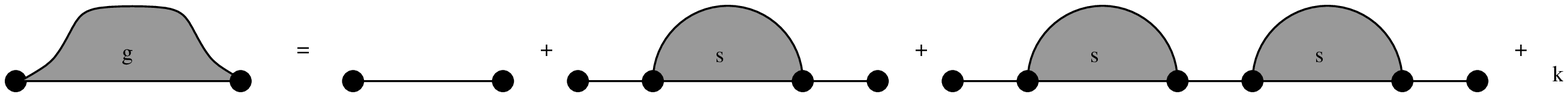}
\end{center}
\begin{center}
\psfrag{s}{{\small ${\bf \Sigma}_*$}}
\psfrag{g}{{\small $\bg_*$}}
\psfrag{k}{{\small $\dots$}}
\includegraphics[width=13cm]{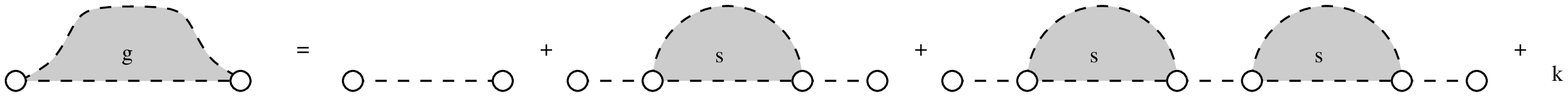}
\end{center}
The reason why it is useful to introduce
the one-line irreducible sums ${\bf \Sigma}(z)$ 
and ${\bf \Sigma}_*(z)$ is that in the large $N$ limit
one can write down two additional, 
independent equations which relate ${\bf \Sigma}(z)$ 
and ${\bf \Sigma}_*(z)$ to ${\bg}(z)$ and ${\bg}_*(z)$.
These are the following
Dyson-Schwinger relations: 
\begin{equation}
\psfrag{g}{${\bg}$}
\psfrag{s}{{${\bf \Sigma_*}$}}
\psfrag{=}{$=$}
\includegraphics[width=5cm]{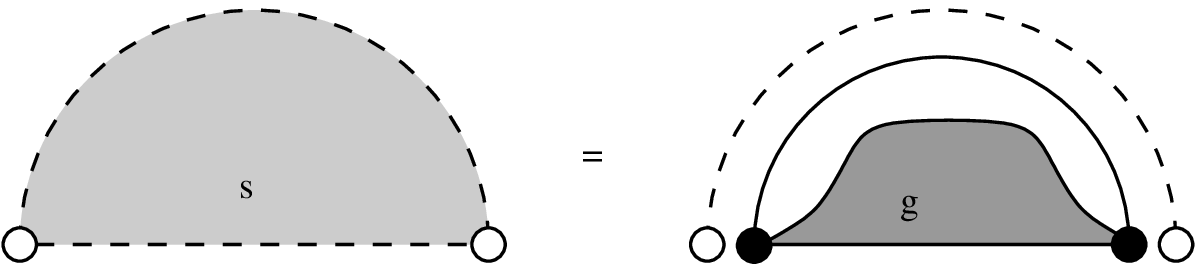} \qquad , \qquad
\psfrag{g}{{$\bg_*$}}
\psfrag{=}{$=$}
\psfrag{s}{{${\bf \Sigma}$}}
\includegraphics[width=5cm]{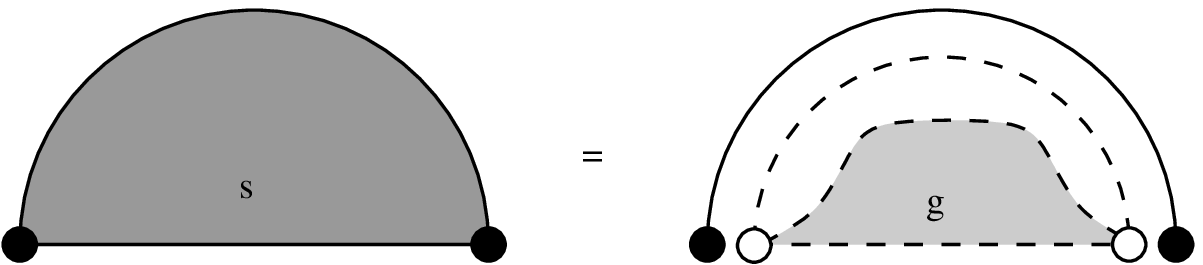}
\end{equation}
which follow from the observation that any planar diagram 
in the sum ${\bg}(z)$ or ${\bg}_*(z)$ can be transformed 
into a one-line irreducible planar diagram by adding an arc 
joining its external points.

In this way one obtains a closed set of equations: one has
four equations for four unknown matrices 
${\bg}$, ${\bf \Sigma}$, $\bg_*$ and ${\bf \Sigma}_*$: 
\begin{eqnarray}
{\bg}(z) &=& \frac{1}{z - {\bf \Sigma}(z)} \label{e1} \\
{\bg}_*(z) &=& \frac{1}{T - {\bf \Sigma}_*(z)} \label{e2}\\
{\bf \Sigma}(z) &=& {\bC}{\rm Tr}\big[ {\bg}_*(z) \big] \label{e3}\\
{\bf \Sigma}_*(z) &=& \mathbbm{1}_T{\rm Tr}\big[{\bg}(z){\bC}\big] \label{e4}
\end{eqnarray}
One can eliminate ${\bf \Sigma}$, $\bg_*$ and ${\bf \Sigma}_*$ 
and solve it for ${\bg}(z)$ as a function of $\bC$. 

It is convenient to write the result using a
variable $Z$ defined by 
\begin{equation}
z {\bg}(z) =Z {\bG}(Z) 
\label{zGzG}
\end{equation}
where ${\bG}(Z)$ is given by equation (\ref{GC}). This amounts to 
choosing $Z$ as
\begin{equation}
Z = \frac{z}{{\rm Tr}{\bg}_*(z)} \ .
\end{equation}
as follows from (\ref{e1}) and (\ref{e3}).
The equations (\ref{e2}) and (\ref{e4}) complete 
the solution with the result
\begin{equation}
Z = \frac{z}{ 1 + r\big(-1+ \frac{1}{N}{\rm Tr} [z {\bg}(z)]\big)} \ .
\label{ztz}
\end{equation}
The equations (\ref{zGzG}) and (\ref{ztz}) relate the resolvent
${\bg}(z)$ calculated in a series of
independent experiments to the genuine resolvent ${\bG}(z)$
for the system. 

In the limit of infinitely many measurements 
$T=\infty$, and $N$ fixed, which corresponds to $r = 0$,
one obtains $Z = z$ and hence ${\bg}(z) = {\bG}(z)$.
As expected, in this limit, spectral densities $\rho(\lambda)$
and $\rho_0(\Lambda)$ are identical.

\end{document}